\documentclass[]{article}
\usepackage{emulateapj}
\usepackage{graphicx}

\advance \voffset by -0.50cm\relax

\def\msun{{\rm ~M}_{\odot}}

\def\mpy{{\rm ~M}_{\odot} {\rm ~yr}^{-1}}

\begin{document}

\title{On the rarity of double black hole binaries: consequences for
gravitational-wave detection}

 \author{Krzysztof Belczynski\altaffilmark{1,2}, 
         Ronald E.\ Taam\altaffilmark{3},
         Vassiliki Kalogera\altaffilmark{3}, 
         Frederic A.\ Rasio \altaffilmark{3},
         Tomasz Bulik\altaffilmark{4,5}}

 \affil{
     $^{1}$ New Mexico State University, Dept of Astronomy,
            1320 Frenger Mall, Las Cruces, NM 88003\\
     $^{2}$ Tombaugh Fellow\\
     $^{3}$ Northwestern University, Dept of Physics \& Astronomy,
           2145 Sheridan Rd, Evanston, IL 60208\\
     $^{4}$ Astronomical Observatory, Warsaw University, Aleje Ujazdowskie
            4, 00478 Warsaw, Poland\\  
     $^{5}$ Nicolaus Copernicus Astronomical Center,
            Bartycka 18, 00-716 Warszawa, Poland;\\
     kbelczyn@nmsu.edu, r-taam, vicky, rasio@northwestern.edu,
     bulik@camk.edu.pl}

 \begin{abstract}
Double black hole binaries are among the most important sources of gravitational
radiation for ground-based detectors such as {\em LIGO} or {\em VIRGO}. Even
if formed with lower efficiency than double neutron star binaries, they
could dominate the predicted detection rates, since
black holes are more massive than neutron stars and therefore could be
detected at greater distances. Here we discuss an evolutionary 
process that can very significantly limit the formation of close double black hole
binaries: the vast majority of their potential progenitors
undergo a common envelope (CE) phase while the donor, one of the massive
binary components, is evolving through the Hertzsprung gap. Our latest
theoretical understanding of the CE process suggests that
this will probably lead to a merger, inhibiting double black hole
formation. Barring uncertainties in the physics of CE evolution, 
we use population synthesis calculations, and find that the corresponding
reduction in the merger rate of double black holes formed in galactic fields is so
great (by $\sim$ 500) that their contribution to inspiral detection rates for
ground-based detectors could become relatively small ($\sim$ 1 in 10) compared to
double neutron star binaries.  A similar process also reduces the merger rates
for double neutron stars, by factor of $\sim 5$, eliminating most of the 
previously predicted ultracompact NS-NS systems. 
Our predicted detection rates for Advanced LIGO are
now much lower for double black holes ($\sim 2$ yr$^{-1}$),
but are still quite high for double neutron stars ($\sim 20$ yr$^{-1}$). If double
black holes were found to be dominant in the detected
inspiral signals, this could indicate that they mainly originate from dense 
star clusters (not included here) or that our theoretical understanding of the CE 
phase requires significant revision.
 \end{abstract}


\section{Introduction}

A number of ground-based gravitational wave detectors are already 
in operation (TAMA, GEO, LIGO) and some are approaching an operational phase
(VIRGO).
These instruments have provided the first upper limits on signals for
some
potential sources of gravitational radiation (Abbott et al.\ 2005a,b,c).
In this work we discuss the likelihood for detecting the gravitational
wave signature of double compact object mergers.
The most promising candidates
include double neutron stars (NS-NS), double black holes (BH-BH) and mixed
systems containing a black hole and a neutron star (BH-NS).
Only NS-NS binaries have been so far discovered in the Galaxy electromagnetically, 
and therefore only for this population are there observational estimates of merger 
rates that can be translated into detection rates for a given detector 
(see Kalogera et al.\ 2004 and Kim, Kalogera, \& Lorimer 2006, for the most 
recent estimates).
The theoretical studies have been carried out most often via population 
synthesis methods, that allow for a self-consistent evolution of massive stars leading
to the formation of all three populations of compact objects.  The
early work was conducted by a number of groups (e.g., Lipunov, Postnov \& Prokhorov
1997; Bethe \& Brown 1998; De Donder \& Vanbeveren 1998; Bloom, Sigurdsson
\& Pols 1999; Fryer, Woosley \& Hartmann 1999; Belczynski \& Bulik 1999; 
Nelemans, Yungelson \& Portegies Zwart 2001) and different results were summarized 
and discussed in Belczynski, Kalogera \& Bulik (2002a: BKB02).

After our initial work (BKB02), our group has been working intensively for several 
years to understand issues involved in the formation of double compact objects and 
to identify the most important (and usually uncertain) processes involved in the 
evolution of massive stars. As a result, we have created a much refined and updated 
population synthesis model (Belczynski et al.\ 2007). This model is now used to 
calculate the merger rates of double compact objects, and to reexamine the chances 
for detection of gravitational waves from double compact object inspirals and mergers.
New results, in the context of gravitational waves and double compact object mergers, 
have been recently reported in O'Shaughnessy, Kalogera \& Belczynski (2006a) and  
O'Shaughnessy et al.\ (2006b).  The results of the above studies are complementary 
to those reported here: O'Shaughnessy et al.\ (2006b) predominantly conducted a broad 
study of the parameter space and applied a number of empirical rate constraints on
population synthesis models; here we discuss the specifics of binary evolution leading 
to the formation of double compact objects, we highlight the importance of CE 
evolution in the Hertzsprung gap for the inspiral of double compact objects, and we 
present updated detection rates in the context of our input physics. In this study we 
focus, in particular, on BH-BH binaries as the results have changed significantly for 
this population and are of great importance for LIGO. A thorough discussion of changes 
(not crucial in the context of GR sources) for NS-NS and BH-NS populations will be 
presented in a forthcoming paper in the context of the progenitors of short-hard 
gamma-ray bursts (GRBs) in Belczynski et al.\ (2007).

\section{Model}

Binary population synthesis is used to calculate the merger rates and properties of 
double compact objects. The formation of double compact objects is modeled via binary 
evolutionary processes that take place without considering the effects of stellar 
dynamical processes associated with the formation of such systems in globular 
clusters. The formation of binary compact objects in dense stellar environments is 
currently under intense investigation (see Portegies Zwart \& McMillan 2000; 
O'Leary et al.\ 2006).

Our population synthesis code, {\tt StarTrack},  was initially developed for the 
study of double compact object mergers in the context of GRB progenitors (Belczynski 
et al. 2002b) and gravitational-wave inspiral sources (BKB02). In recent years {\tt 
StarTrack} has undergone major updates and revisions in the physical treatment of
various binary evolution phases, and especially mass transfer phases. The new version 
has already been tested and calibrated against observations and detailed binary mass 
transfer calculations (Belczynski et al.\ 2007), and has been used in various 
applications (e.g., Belczynski \& Taam 2004; Belczynski et al.\ 2004; Belczynski,
Bulik \& Ruiter 2005). The physics updates that are most important for compact
object formation and evolution include: a full numerical approach to
orbital evolution due to tidal interactions, calibrated 
using high mass X-ray binaries and open cluster observations, a
detailed treatment of mass transfer episodes fully calibrated against
detailed calculations with a stellar evolution code, updated stellar
winds for massive stars, and the latest determination of the natal
kick velocity distribution for neutron stars (Hobbs et al.\ 2005). For helium star 
evolution, which is of a crucial importance for the formation of double neutron star 
binaries (e.g., Ivanova et al.\ 2003;  Dewi \& Pols 2003), we have applied a 
treatment matching closely the results of detailed evolutionary calculations.
If the helium star fills its Roche lobe, the systems are examined for the 
potential development of a dynamical instability, in which case they 
are evolved through a CE phase, otherwise a highly non-conservative mass
transfer issues.  
We treat CE events using the energy formalism (Webbink 1984),
where the binding energy of the envelope is determined from the set 
of He star models calculated with the detailed evolutionary code by 
Ivanova et al.\ (2003). The progenitor evolution and the Roche lobe overflow episodes
are now followed in much greater detail. We note significant
differences from our earlier studies. For a detailed description of the
revised code we refer the reader to Belczynski et al.\ (2007).

The most important change in the context of double compact object
formation reflects the treatment of the dynamically unstable
mass transfer and evolution into the CE phase. Previously (e.g.,
BKB02, and earlier work) we have allowed for the possibility of binary
survival when the CE was initiated by a star crossing the Hertzsprung gap 
(HG). Once the system evolved into the CE, it was determined whether a
system emerges as a tight post-CE binary or produces a single, rapidly 
rotating star as a result of a binary merger. However, a HG star  
does not have a clear entropy jump at the core-envelope transition 
(Ivanova \& Taam 2004).  Therefore, once a companion is engulfed within 
a HG star there is no clear boundary where the inspiral ceases and consequently a 
merger is expected (see e.g., Taam \& Sandquist 2000).  In this case, 
the HG star is insufficiently evolved such that the systems merge even 
though it may be energetically possible to unbind the envelope. It is 
expected that the orbital decay timescale is shorter than the mass loss 
timescale. This is a direct result of the absence of steep density gradients 
above the nuclear burning shells in this evolutionary stage, where a 
non negligible amount of mass lies above the core facilitating 
rapid decay of the orbit. In the current {\tt StarTrack} modeling, we assume that any CE
involving a HG donor eventually leads to a binary component merger and
to the formation of a single object. Since many potential double compact 
object progenitors evolve through the CE phase (see BKB02 and their 
Table~3), we have reexamined the double compact object populations in view 
of these mergers expected in CEs with HG donors. 
In particular it is found that close BH-BH systems are affected the most, 
as the majority of their potential progenitors evolve through the CE
phase with a hydrogen-rich HG donor. It is also found that about half of 
the progenitors of NS-NS systems evolve through a CE phase with helium-rich  
HG donors.  These systems were allowed in our previous work to form very close, 
ultracompact, NS-NS systems. With this study, we now recognized,
that these systems are not likely to survive the CE phase, and we treat them
as mergers (inhibiting significantly ultracompact NS-NS formation) in our 
reference model. 
However, we also present alternative models in which we apply the CE energy 
formulation even in the case of HG donors to test this uncertain phase of 
binary evolution.

Another important change reflects the treatment of accretion onto compact
objects in CE phases. In some cases the CE phase may be initiated by a star
that has evolved beyond the HG (e.g., red giant or massive core-helium burning star). 
Up to now, the full Bondi-Hoyle accretion rate onto compact object
(for numerical details see BKB02) was utilized. On average, it resulted in 
the accretion of $\sim 0.4 \msun$ onto NSs, and up to several solar masses onto BHs.      
However, we have recognized that if such efficient accretion is allowed,
the predicted masses of NSs in NS-NS systems will be always too high as compared
with masses for the observed Galactic systems ($\simeq 1.35 \msun$, Thorsett \& 
Chakrabarty 1999). Motivated by this observation and by multi dimensional hydrodynamical 
calculations that favor mass accretion rates lower than Bondi-Hoyle (e.g., Ruffert 1999), 
for our reference model we allow only for smaller amounts of mass gain ($0.05-0.1 \msun$) 
in CE phases. The specific values were chosen such as to be sufficient to
(mildly) recycle a NS (e.g., Zdunik, Haensel \& Gourgoulhon 2002; Jacoby et al. 2005). 
Since such small amounts of accreted mass may not be very realistic in the 
case of BHs, we also present a model with full Bondi-Hoyle accretion.

\section{Results}

Evolution leading to the formation of a BH-BH binary involves a number
of stages that allow two massive stars to evolve to close proximity and to form a 
bound BH-BH system. Here, we are interested only in those binaries that form BH-BH 
systems on very tight orbits, so the timescale for inspiral due to gravitational 
radiation is shorter than 10\,Gyr. Only these BH-BH systems can merge and contribute 
to gravitational wave detection rates (at least in galaxies with star formation 
history similar to the Milky Way). There are a number of processes that can inhibit 
the formation of BH-BH systems.  The main factors involve mergers resulting from 
mass transfer episodes (single merger remnant is formed), binary disruption upon 
the formation of a BH due to mass loss and the potential natal kick that a BH may 
receive. However, it is important to realize that these same processes (e.g., 
adequately placed natal kick) can allow in some circumstances the formation of 
close BH-BH systems. Here we present a set of population synthesis models that allow 
to investigate the effects of those binary evolution elements most important for 
BH-BH formation. Model A is our reference model with input physics described in 
detail in Belczynski et al.\ (2007) and the most important new elements relevant to 
double compact object formation are emphasized in \S\,2. Other models are examined 
with only one input parameter changed compared to the reference model. In model B, 
we allow for full hyper-critical (Bondi-Hoyle) accretion onto compact objects 
(see BKB02 for the details of the implementation). In models A and B, CE events with 
both hydrogen-rich and helium-rich HG donor are assumed to lead to mergers (as 
explained in \S\,2). However, in model C, we allow for possible survival of the 
binaries through such CE events and apply the standard CE energy formulation; this 
model is the one closest to our reference/standard models is past synthesis studies 
by our group (e.g., BKB02).   

The major formation channels for close BH-BH binaries are illustrated in 
Table~\ref{channels}. In models A and B BH-BH formation typically does not involve CE 
phases, but consists of non-conservative mass transfer episodes and supernovae/core 
collapse events in which BHs are formed (channels BHBH:01,02,04). Only in one channel 
(BHBH:03) is CE evolution encountered, but in this case it involves evolved donors 
(during core helium burning) and not HG stars. In model C, most BH-BH systems form via 
evolutionary channels involving CE phases, and, in most cases the donors are HG stars 
(BHBH:05,06). Channels not involving CE phases with HG donors are also found in model C, 
but they are very rare and are included, among other inefficient sequences, in channel 
BHBH:07.

The calculated Galactic merger rates for double compact objects
are presented in Table~\ref{rates}.  The rates are obtained for two different calibrations. 
First, we have calibrated our results based on the star formation rate. We have adopted a 
continuous star formation rate in the Galaxy of $3.5 \mpy$ (O'Shaughnessy et al. 2006b)  
lasting for 10 Gyr. In addition, a  combined rate of SN II and SN Ib/c estimated for a 
Milky Way type galaxy of $2 \times 10^{-2}$ yr$^{-1}$ is used to obtain an 
alternative calibration (Cappellaro, Evans \& Turatto 1999 for a Galactic blue
luminosity of $L_{\rm B} = 2 \times 10^{10} L_\odot$). The supernova-based calibration 
results in merger rates about 1.5 times higher than those for the star formation calibration.  
For models A and B we find that NS-NS merger rates are $7.6-19$ Myr$^{-1}$, consistent with 
the most recent empirical estimates ($3-190$ Myr$^{-1}$; see Kim, Kalogera, \& Lorimer 2006). 
However, they are somewhat lower (as expected) when compared with our earlier estimates 
($\sim 50$ Myr$^{-1}$; see the standard model in BKB02). The rates for binaries containing 
black holes are significantly lower in models A and B; $0.07-0.14$ Myr$^{-1}$ and 
$0.01-0.03$ Myr$^{-1}$ for BH-NS and BH-BH systems, respectively. In particular, these rates 
are much lower than previously reported ($\sim 8$ Myr$^{-1}$ and $\sim 26$ Myr$^{-1}$ for 
BH-NS and BH-BH systems in the standard model in BKB02). The rates for model C where we allow 
for survival in CE with HG donors, although such a survival is considered unlikely (see \S\,2), 
are closest to our previous calculations. The differences (factor of $\sim 2$) with respect to 
rates presented by BKB02 are a consequence of a number of improvements and updates in the 
{\tt StarTrack} code (see \S\,2 and also Belczynski et al.\ 2007).  
In this model C again the Galactic merger rates of NS-NS are the highest
($\sim 90$ Myr$^{-1}$), however rates for BH-NS ($\sim 4$ Myr$^{-1}$) and
BH-BH ($\sim 10$ Myr$^{-1}$) mergers are of the same order of magnitude, in contrast to 
models A and B. We note that the large decrease (factors of $\sim 300-800$) in BH-BH merger 
rates from model C to models A and B, is due to the expectation that CE events with HG donors 
lead to mergers.

In Table~\ref{ligo} we show the expected detection rates for the initial
(current stage) and the advanced (expected in 2014) LIGO. The
detection rates were obtained from the predicted Galactic inspiral rates,
${\cal R_{\rm MW}}$ (see Table~\ref{rates}), with  
\begin{equation}
{\cal R_{\rm LIGO}} = \rho_{\rm gal} {4 \pi \over 3} d_0^3 {\cal M}_{\rm
dis}
{\cal R_{\rm MW}}
\label{ligrat}
\end{equation}
where $\rho_{\rm gal}=0.01$ Mpc$^{-3}$ is the number density of Milky
Way-type galaxies that approximates the mass distribution within the 
LIGO
distance range $d=d_0 ({{\cal M}_{\rm c,dco} / {\cal M}_{\rm
c,nsns}})^{5/6}$,
with $d_0=18.4,\ 300$ Mpc for the initial and advanced LIGO respectively. 
The distance range estimates were obtained for a binary with two
$1.4 \msun$ neutron stars with chirp mass of ${\cal M}_{\rm c,nsns}=1.2 \msun$,
and we rescale them for our populations of double compact objects for given 
chirp masses ${\cal M}_{\rm c,dco}$.  The scaling factor is obtained from
\begin{equation}
{\cal M}_{\rm dis} = \left< \left( {{\cal M}_{\rm c,dco} /
{\cal M}_{\rm c,nsns}} \right)^{15/6}. \right>
\label{dscale}
\end{equation}
Note that we first calculate the cube of $({{\cal M}_{\rm
c,dco}/
{\cal M}_{\rm c,nsns}})^{5/6}$ and then take an average over all
double compact objects within a given group (e.g. BH-BH).  
This calculation assumes Euclidean space geometry and a constant
star formation rate, while the cosmological effects, relevant for BH-BH
merger advanced LIGO rate, are discussed by Bulik, Belczynski \& Rudak
(2004).  The scaling factors ${\cal M}_{\rm dis}$ for different groups of
double
compact objects are given in Table~\ref{rates}. The specific values for
$\rho_{\rm gal}$, $d_0$ and ${\cal M}_{\rm c,nsns}$ were adopted from
O'Shaughnessy et al.\ (2006b).

For initial LIGO we find that the rates are too small for 
detection in agreement with predictions in our earlier work, and
now confirmed with the new models. 
For advanced LIGO, on the other hand, we expect quite significant detection rates. 
For physical models (A and B) the detection rate is dominated by NS-NS events 
($\sim 20$ yr$^{-1}$), with a small contribution of BH-BH inspirals ($\sim 2$ yr$^{-1}$), 
and an even smaller contribution of BH-NS mergers ($\sim 1$ yr$^{-1}$). This is
a qualitatively new result, as earlier work (e.g., BKB02) expected BH-BH binaries 
to dominate the detection rates as demonstrated in model C.
In particular, inspiral signals were expected to be detected at much
higher rates ($\sim$ hundreds a year; see model C), dominated by populous
BH-BH mergers, as compared to new smaller rates ($\sim$ tens a year; see
models A and B) obtained for models with a dominant NS-NS population.

\begin{figure}
\includegraphics[width=1.1\columnwidth]{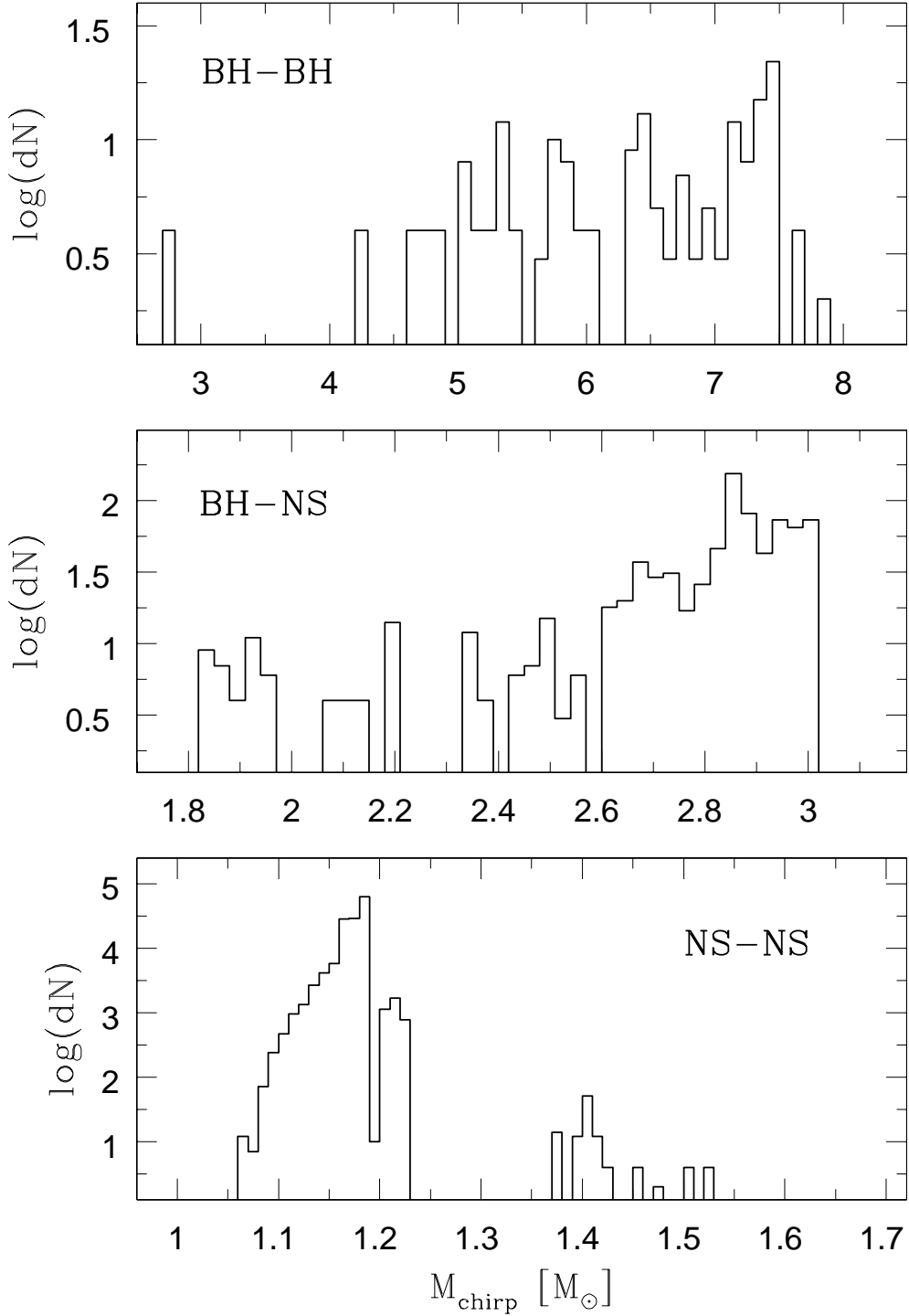}
\caption{Chirp mass distribution for different populations of double compact
objects (reference model: A). Chirp masses are of the order of $\sim 1.1-1.2 \msun$
for NS-NS, $\sim 2.5-3 \msun$ for BH-NS, and $\sim 5-8 \msun$ for BH-BH
mergers. Note the range changes from panel to panel.  
}
\label{chipA}
\end{figure}

\begin{figure}
\includegraphics[width=1.1\columnwidth]{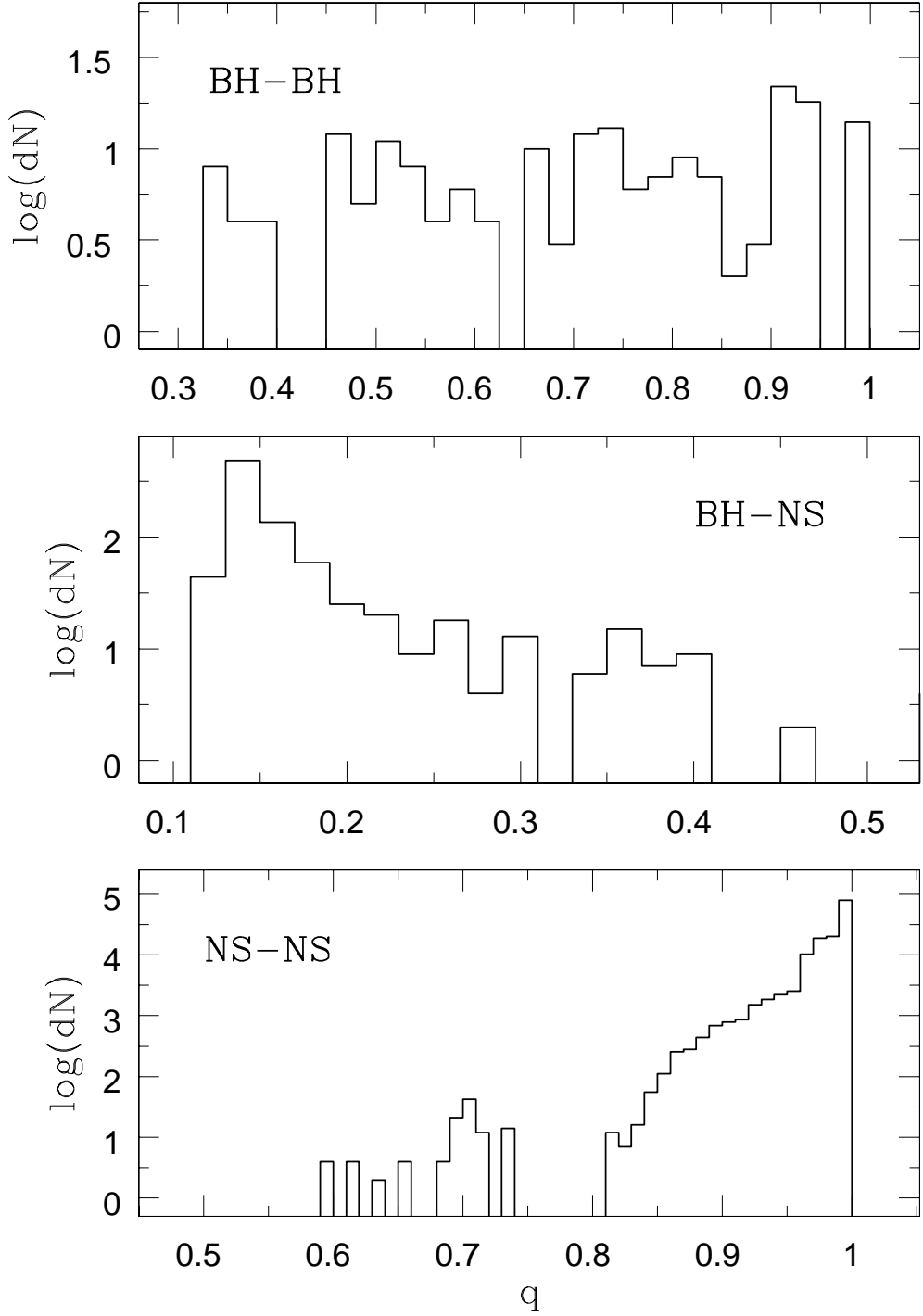}
\caption{Mass ratio distribution for different populations of double compact
object mergers (reference model: A). Note that majority of NS-NS mergers are found 
with very high mass ratios ($q>0.9$), BH-NS mergers with low mass ratios
($q<0.4$), while the distribution is rather flat for BH-BH mergers with 50\% 
of them characterized by high mass ratio ($q>0.74$).
}
\label{qA}
\end{figure}

\begin{figure}
\includegraphics[width=1.1\columnwidth]{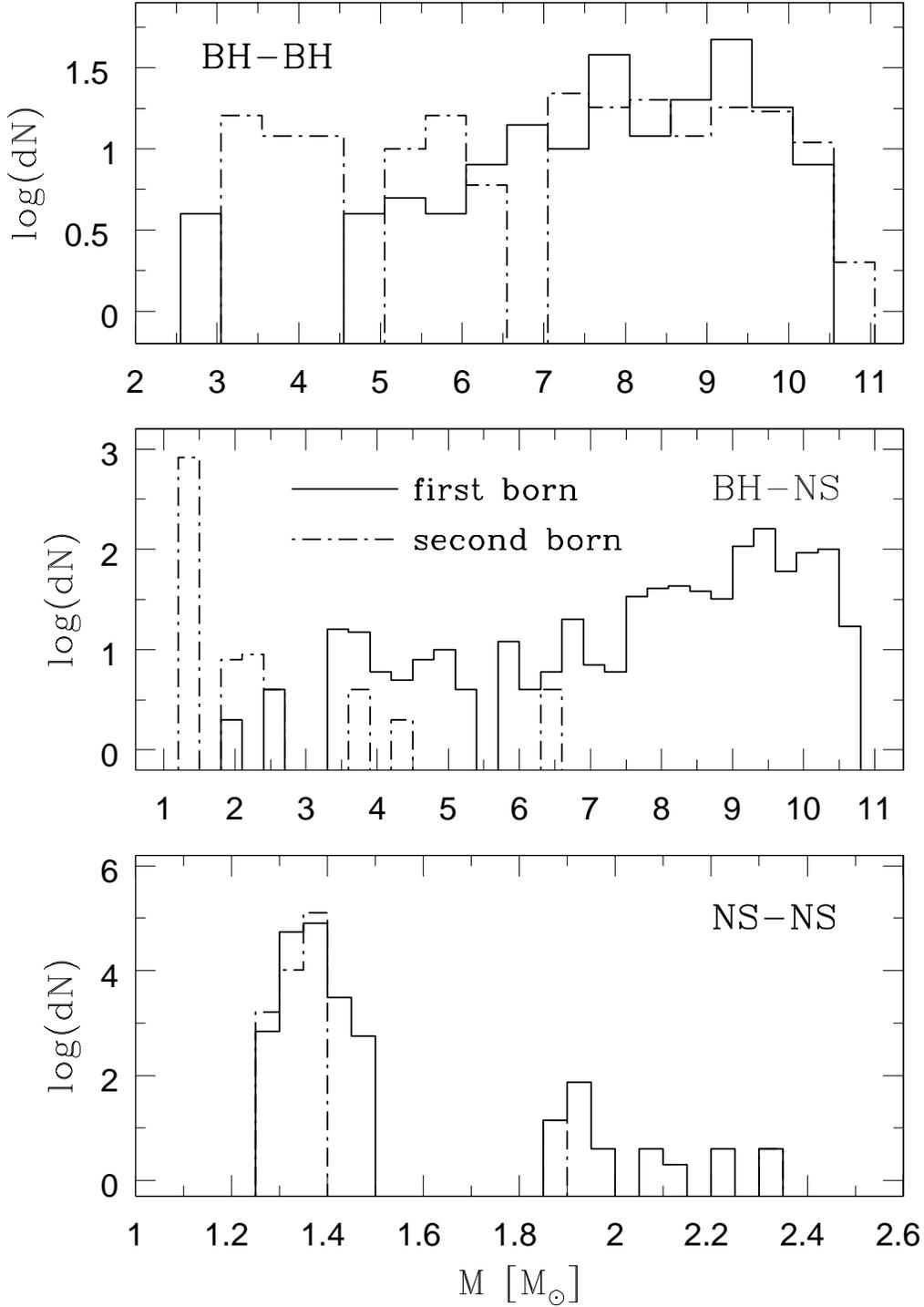}
\caption{Compact object masses for different populations of double compact
object mergers (reference model: A). First- and second-born compact objects
are marked with different lines. Note that NS masses peak at $\sim 1.35
\msun$, while BH are found in wide range of masses $\sim 3-11 \msun$ with
increasing contribution at the higher mass end.  
}
\label{MabA}
\end{figure}

In Figures~\ref{chipA}, ~\ref{qA} and ~\ref{MabA} we show the 
chirp mass, mass ratio, and component mass distributions for double
compact objects in the reference model (model A) respectively. 
The figures include data on double compact objects that have merged in 
10 Gyr. The numbers correspond to the disk population of a Milky Way
type galaxy with a constant star formation rate of the order of $3.5 \mpy$.    
As expected chirp masses are highest for BH-BH systems ($\sim 5-8
\msun$), intermediate for BH-NS mergers ($\sim 2.5-3 \msun$) and  lowest
for the lightest NS-NS systems ($\sim 1.1-1.2 \msun$). The mass ratio distribution 
is rather flat for BH-BH mergers; however due to the fact that it starts at
rather high value ($q \sim 0.5$), 50\% of the systems are found with a
BH of similar mass ($q > 0.74$). Mass ratios for BH-NS are generally low 
($q<0.2$) with the BH being the much more massive component in a given system. 
For most NS-NS mergers the mass ratios are rather high ($q > 0.9$) since usually 
two NSs have very similar masses. Component masses in BH-BH systems are found in a 
wide range ($3-11 \msun$), 
although most BHs in the merging population are characterized by rather high
mass ($\sim 7-10 \msun$). For NS-NS systems the majority of component masses are
confined to a narrow range ($1.1-1.5 \msun$), with a small contribution of
heavy NSs ($\sim 2 \msun$). For mixed BH-NS mergers, the masses are high for
BH and low for NS, explaining the rather extreme mass ratio distribution. Note 
that for some systems the first born compact object is a NS and not a BH. This
is due to a mass ratio reversal occurring during the mass transfer phase, leading 
to mass loss from the primary (NS formation) and the rejuvenation of the secondary 
(BH formation). 

\begin{figure}
\includegraphics[width=1.1\columnwidth]{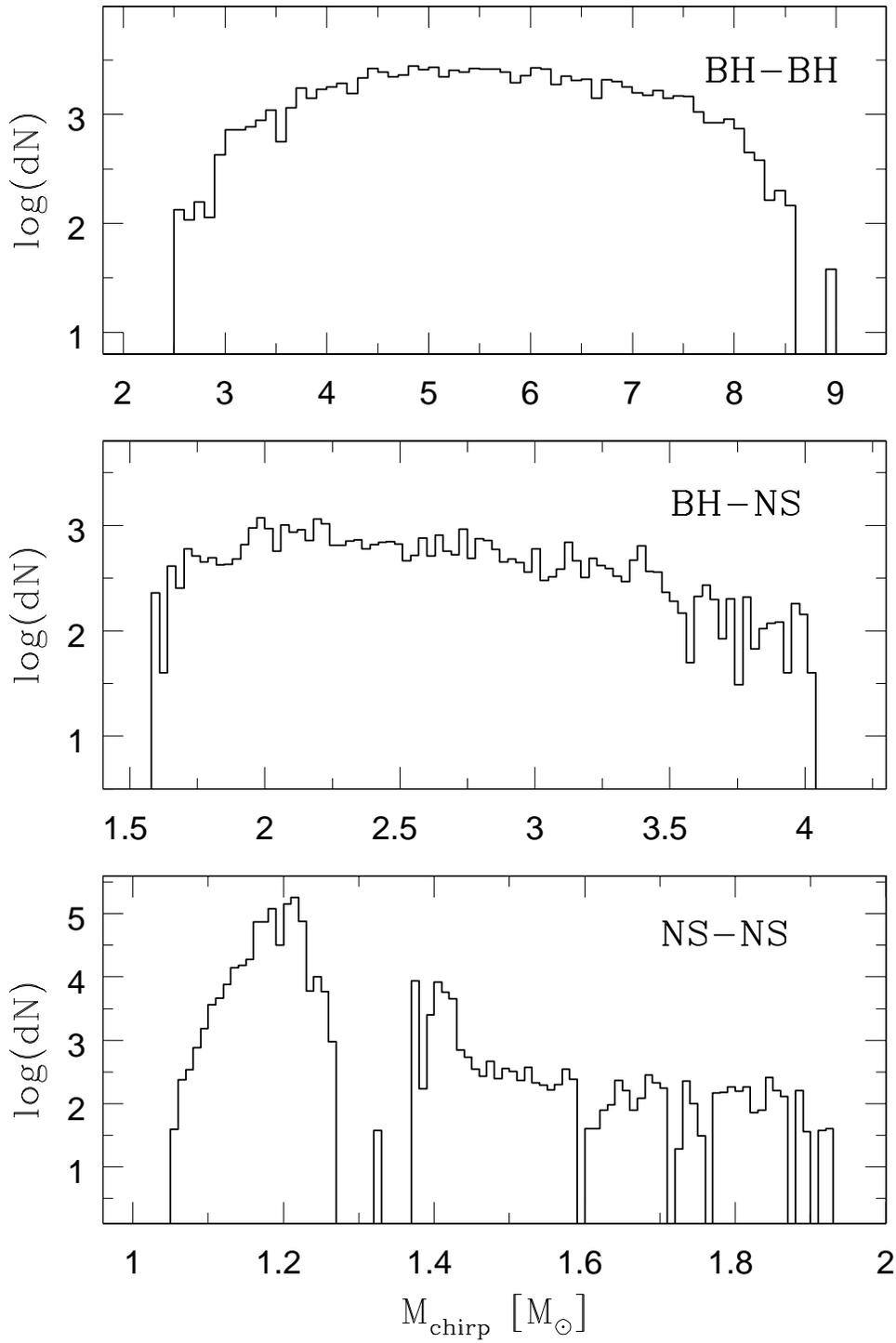}
\caption{Chirp mass distribution for different populations of double compact
objects (model C). 
}
\label{chipC}
\end{figure}

\begin{figure}
\includegraphics[width=1.1\columnwidth]{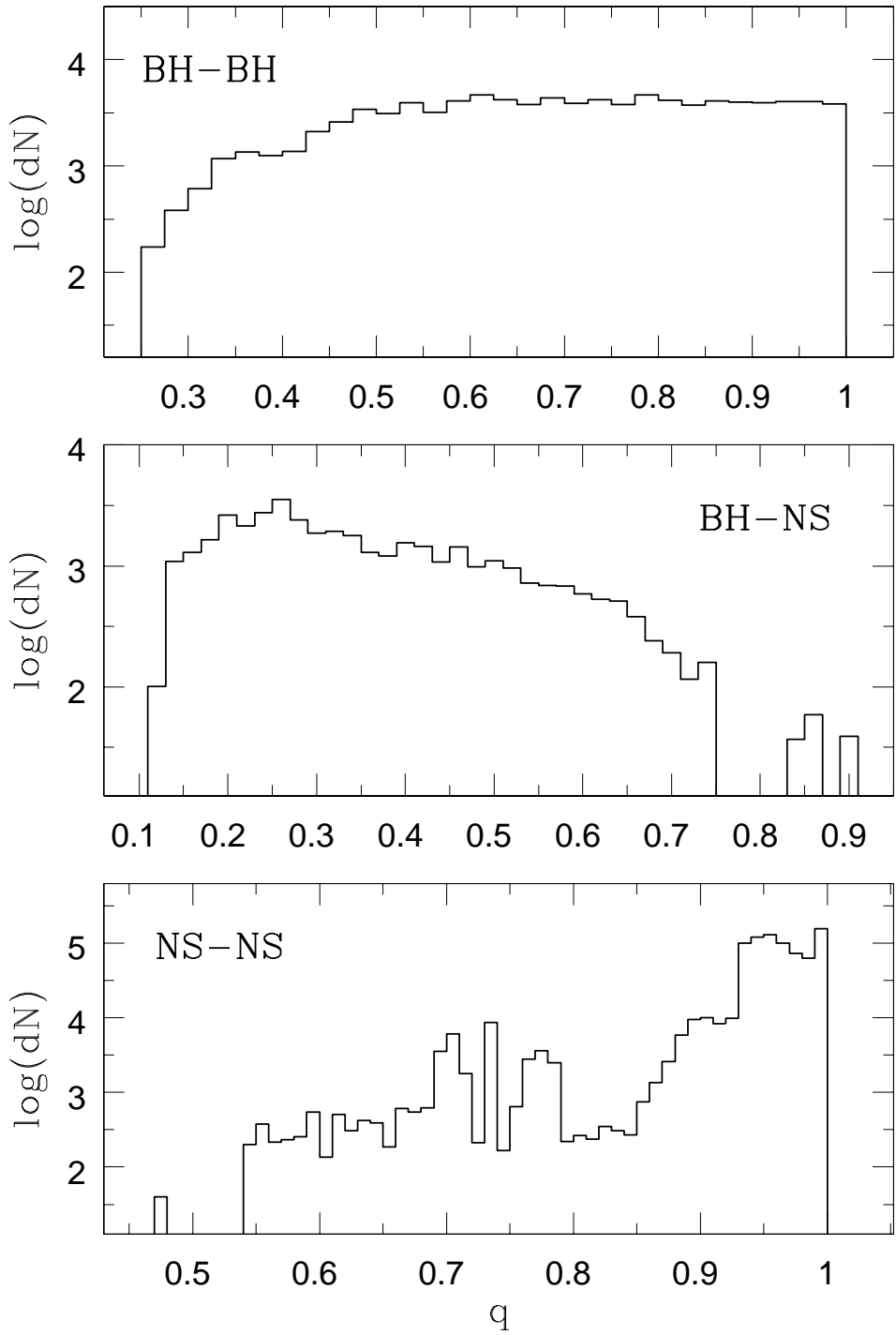}
\caption{Mass ratio distribution for different populations of double compact
object mergers (model C). 
}
\label{qC}
\end{figure}

\begin{figure}
\includegraphics[width=1.1\columnwidth]{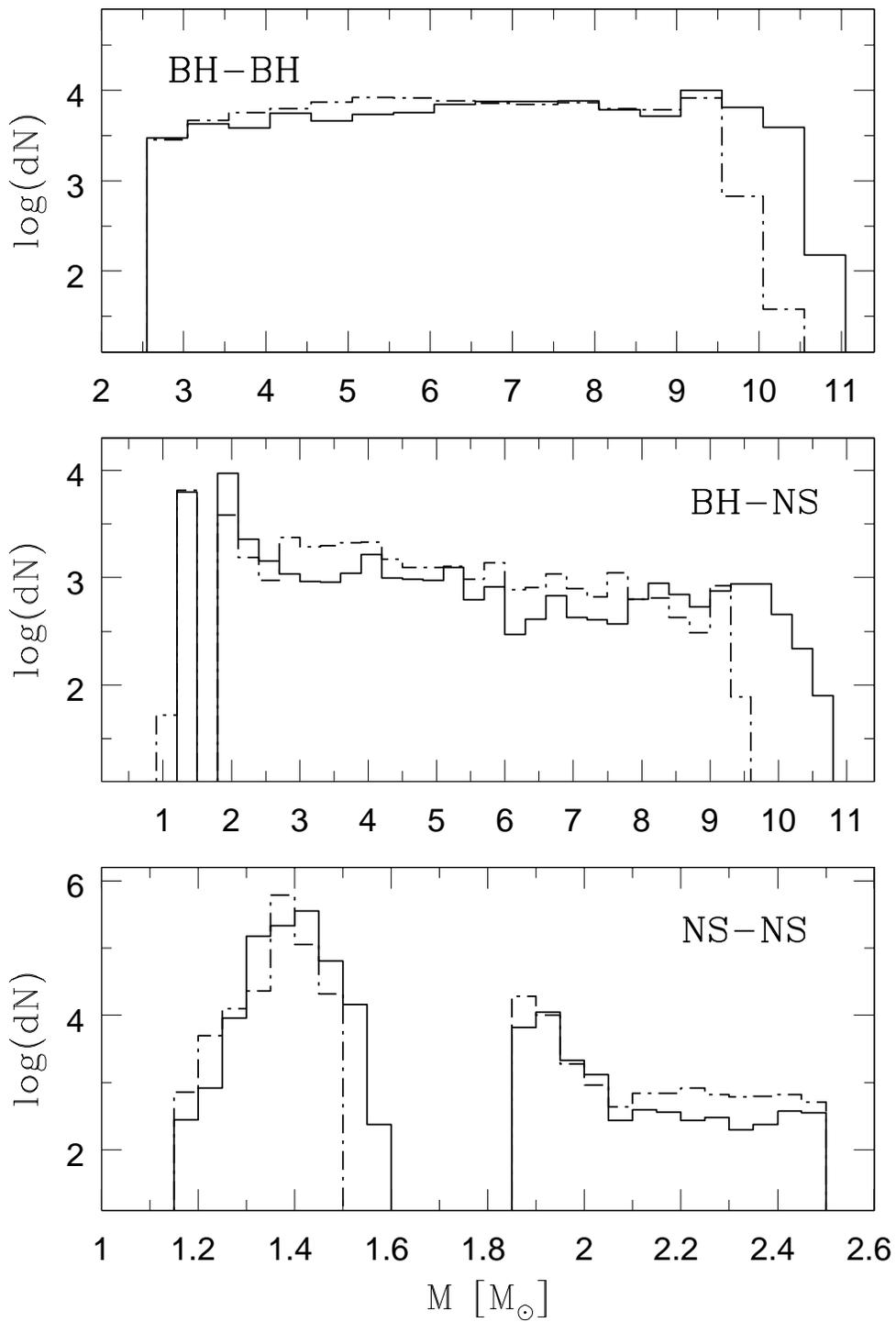}
\caption{Compact object masses for different populations of double compact
object mergers (model C). Notation same as in Fig.~\ref{MabA}.
}
\label{MabC}
\end{figure}

In Figures~\ref{chipC}, ~\ref{qC} and ~\ref{MabC} we present similar
distributions for model C. Since formation of close BH-BH
and BH-NS systems is much more efficient (although probably unphysical) 
in this model, a much higher number of these systems is evident in the 
distributions. As compared to our new reference model, chirp masses in the model C 
for BH-BH ($\sim 3-8 \msun$), BH-NS ($\sim 1.5-4 \msun$) and NS-NS ($\sim 1.1-1.8 \msun$)  
are distributed more evenly and over a wider range. This is due to the
fact that many more primordial binaries (with different initial properties) are 
allowed to contribute in this model to the formation of double compact objects. 
Mass ratios for BH-BH are found evenly distributed over $q>0.5$ with a tail
in the distribution extending to slightly lower values ($q \sim 0.3$). For BH-NS
systems the majority of mass ratios are small ($q \sim 0.1-0.4 $), while for
NS-NS systems mass ratios are rather large ($q>0.9$) similar to model A. 
Individual component
masses are distributed much more evenly than in the reference model for BH-BH
mergers. For BH-NS systems, component masses tend to be much smaller in model 
C ($\sim 2-4 \msun$) than in the reference model ($\sim 8-11 \msun$). 
For NS-NS systems, the component masses are similar to the reference model,
with an increased contribution of high mass NSs.

We note that model (B) with full Bondi-Hoyle accretion is inconsistent with observed NS
masses in Galactic NS-NS binaries. The masses of the (second formed) NS stars in this 
model are found between $\sim 1.4-1.8 \msun$, while observed masses are
estimated to be $\sim 1.35 \msun$ (for more details see Belczynski et al.
2007, in preparation). For BH-BH systems only $\sim 13\%$ of
these evolve through potential CE accretion (see Table~\ref{channels}). This 
may increase the BH masses by a few solar masses, thereby shifting the mass 
distributions to slightly higher values.

\section{Discussion}

We have presented new results for double compact object mergers with
special emphasis on BH-BH systems. We identify an evolutionary process that  
can drastically decrease the BH-BH merger rates predicted from population 
synthesis models of binaries in galactic fields: mergers of binaries in the 
first CE phase where the donor star is in the HG, which corresponds to the 
majority of potential BH-BH progenitors considered in earlier work. 
The new mergers rates for these systems are significantly lower ($\sim 500$) than 
previously predicted (e.g., model C here, and BKB02). As a consequence BH-BH mergers 
could become only a small contributor ($\sim$ 1 in 10 detections, see models A and B; 
see also O'Shaughnessy et al. 2006b\footnote{O'Shaughnessy et al. (2006b) in a study 
that focused on deriving empirically constrained merger rates explored a very large 
model parameter space; although, naturally, the BH binary merger rates are not always 
as low as for the two models A and B presented here, they too find that LIGO 
detection rates are expected to be dominated by NS-NS mergers, without however 
identifying specific evolutionary phases responsible for this effect.}) 
to inspiral detection rates for ground based detectors like LIGO, in contrast to 
past predictions in which the BH-BH inspiral strongly dominated the detection rates.

If our understanding of the stellar structure of HG stars or the physics of ejection 
of a CE is incomplete, and massive systems with HG donors survive the CE
phase, forming BH-BH systems, it is predicted that BH-BH inspirals will instead dominate
detection rates (see model C). Therefore, once a sizable number of
inspiral signals have been detected (advanced LIGO, see Table~\ref{ligo}),
it may be possible to test this crucial phase of binary evolution leading to
the formation of BH-BH systems. Since the observed chirp mass distribution is a very 
sensitive function of the model of binary evolution, it can lead to further 
constraints on other processes important for the formation of BH-BH systems 
(Bulik \& Belczynski 2003; Bulik et al. 2004). For example, the orientation 
and magnitude of the spin of a compact object is an important factor in the 
search for a gravitational radiation signal from inspiraling binaries since 
it affects both the shape and amplitude of the signal. An initial estimate of 
compact object spins via population synthesis was presented by O'Shaughnessy et al. 
(2005), who identified accretion in the CE phase as a major contributing factor that 
can significantly increase spins of compact objects. These early results need to be 
reassessed since {\em (i)} now only very few BH-BH progenitor systems evolve through 
CE (see Table~\ref{channels}), {\em (ii)} for systems evolving through CE the
so far applied Bondi-Hoyle accretion model may result in overestimates of the
accretion (and spin-up) rates (see \S\,2), {\em (iii)} during the stable mass
transfer phases the spin-up requires recalculation to take account of the possibility 
that the BH can accept mass at rates $\gtrsim 100$ times critical (Eddington) accretion 
rates (e.g., Abramowicz et al. 1988; Ohsuga et al. 2002). A model for BH spin 
evolution is in progress (Belczynski \& Taam 2007, in preparation).     

One additional calculation was performed to examine the effect of CE
treatment on our main result; i.e., the drastic merger rate decrease of BH-BH
binaries. We have applied an alternative prescription for the CE treatment 
involving explicit angular momentum conservation with implicit energy 
conservation (see Nelemans \& Tout 2005). There is only a small change for BH-BH 
rates as CE phase is involved in the formation of only a minor fraction of these 
binaries (see Table~\ref{channels}).

As discussed throughout this study the decrease in merger rates is smaller
for BH-NS and NS-NS binaries as compared with BH-BH systems.  This originates
from the fact that, for these systems, the mass ratios are generally
smaller at the onset of mass transfer, and in many cases dynamical
instability (followed by CE phase) does not develop. Also, the CE phase does 
not often involve a HG donor and the progenitors of BH-NS and NS-NS systems 
are likely to survive. Mixed systems, BH-NS, are rather rare (see also BKB02 
and O'Shaughnessy et al.\ 2006b) and are not predicted to contribute 
significantly to inspiral detection rates (lower by about a factor of 2 
compared to BH-BH binaries in the models presented here; $\sim$ 1 in 20 
detections, see models A and B). In addition, the Galactic merger rates for NS-NS 
systems are modified, but not as significantly as the BH-BH rates. There is a 
decrease (by a factor of $\sim 5$) in merger rates between models A ($\sim 20$ Myr$^{-1}$) 
and model C ($\sim 100$ Myr$^{-1}$). This decrease can be attributed, in approximately even
numbers, to {\em (i)} early evolution with progenitors evolving through CE 
with hydrogen-rich HG massive donor and {\em (ii)} late evolution with
progenitors evolving through CE with HG low-mass helium star donor.   
Phase {\em (ii)} eliminates most of the previously identified ultracompact
NS-NS binaries (e.g., Belczynski \& Kalogera 2001) while phase {\em (i)} eliminates 
some of the wider classical systems. Nevertheless, the NS-NS merger rates from models 
A and B reported here are still consistent with the most recent 
empirical estimates ($\sim 3-190$ Myr$^{-1}$, Kim, Kalogera, \& Lorimer 2006). 
Provided 
that our understanding of CE phases with HG donors is correct, NS-NS mergers are 
predicted to strongly dominate the inspiral signal ($\sim 90\%$ of inspiral events). 
Although the rates are too small to expect a detection at the current initial
LIGO stage ($\sim 0.005$ yr$^{-1}$), they are quite high for Advanced LIGO with
tens of detections expected every year ($\sim 20$ yr$^{-1}$). 

Mergers of compact objects, either NS-NS or BH-NS, are currently the leading 
progenitor model for short-hard GRBs (for a recent review see Nakar 2007). We
have recently presented a study devoted to the connection of these mergers with
short-hard GRBs (Belczynski et al. 2006). In view of the decreased BH-NS rates 
reported here, such mergers may become less favored as potential GRB progenitors 
compared to NS-NS mergers. Although formation of the previously identified 
ultracompact NS-NS population is inhibited, if CE events with HG donors always 
lead to mergers, we still find that a small, but significant, fraction of NS-NS
binaries has merger times shorter than $\sim 100$\,Myr, and such mergers would be 
taking place in star-forming galaxies. We address these and other related issues 
in the forthcoming paper mentioned above Belczynski et al.\ 2007 (in preparation).

Our primary focus in this study has been to examine the important effect CE episodes 
with HG donors have on BH-BH formation and merger rates; we point out this importance 
by comparing a set of only three models that clearly illustrate the important physical 
effects relevant to this issue. Given the small number of models however, the reported 
rates do not fully represent the intrinsic, quantitative uncertainties associated with 
rate predictions from population synthesis calculations (see, e.g., O'Shaughnessy et 
al.\ 2006b where a broader but less specific model exploration has been presented, for 
a discussion of the issue of rate uncertainties).

Last, we emphasize that these estimates do not include the contribution from systems 
formed in dense star clusters, as we have considered only the evolution of field
stars. Following the example of Milky Way, we note that only 1 out of 5 
close NS-NS systems is found in a Galactic globular cluster\footnote{It will be 1 
out of 6, if PSR~J1906+0746 is confirmed as NS-NS binary (Lorimer et al.
2006).} and we do not expect a significant rate increase for NS-NS merger
detection (Phinney 1991). However, it has been pointed out that dynamical interactions 
in dense stellar systems (e.g., globular clusters) can potentially produce close BH-BH systems 
more effectively than in a field population (e.g., Portegies Zwart \& McMillan 2000; 
O'Leary et al. 2006). This is an important issue and predictions depend on the 
number of stars in dense clusters within the reach of ground-based gravitational wave 
detectors, the initial conditions of these clusters, and the interplay between 
single and binary star evolution and dynamical processes. These issues are 
currently under investigation (e.g. O'Leary, O'Shaughnessy \& Rasio in preparation; 
Sadowski et al.\ in preparation; Ivanova et al.\ in preparation), and  once the 
results are available, they should be combined with the field merger rates, 
for a complete description of anticipated inspiral and merger detections with LIGO.

\acknowledgements
We express special thanks to R.O'Shaughnessy for useful discussions and 
help with computationally demanding calculations and an anonymous referee
for several useful comments. 
KB thanks the Northwestern University Theoretical Astrophysics Group for 
its extended, warm hospitality.
We acknowledge partial support through KBN Grant 1P03D02228 and 1P03D00530
(KB, TB), a Packard Foundation Fellowship in Science and Engineering and
NSF grants PHY-0353111 and AST-0449558 (VK) and NSF Grant No. AST-0200876
(RT).

\clearpage

\begin{deluxetable}{lrl}
\tablewidth{350pt}
\tablecaption{Double Black Hole Formation Channels}
\tablehead{ Formation       & Relative   &  \\
            Channel (Model) & Efficiency \tablenotemark{a} & Evolutionary
History
\tablenotemark{b}  }
\startdata

BHBH:01 (A, B)& 50\   \%& NC:a$\rightarrow$b, SN:a, SN:b\\
BHBH:02 (A, B)& 25\   \%& SN:a, SN:b\\
BHBH:03 (A, B)& 13\   \%& SN:a, CE:b$\rightarrow$a, SN:b\\
BHBH:04 (A, B)& 12\   \%& NC:a$\rightarrow$b, SN:a, NC:b$\rightarrow$a, SN:b\\

&&\\

BHBH:05 (C)& 65\ \%& NC:a$\rightarrow$b, SN:a, CE:b$\rightarrow$a, SN:b\\
BHBH:06 (C)& 28\ \%& NC:a$\rightarrow$b, CE:b$\rightarrow$a, SN:a, SN:b\\
BHBH:07 (C)& 7\ \%& all other \\

\enddata
\label{channels}
\tablenotetext{a}{Normalized to the total number of close BH-BH population
in a given model.}
\tablenotetext{b}{Sequences of different evolutionary phases for the primary
(a) and the secondary (b):
non-conservative mass transfer (NC), common envelope (CE),
supernova explosion/core collapse event (SN).
Arrows mark direction of mass transfer episodes.}
\end{deluxetable}

\begin{deluxetable}{lcccl}
\tablewidth{350pt}
\tablecaption{Galactic Merger Rates\tablenotemark{a}\ \ [Myr$^{-1}$]}
\tablehead{Model & NS-NS & BH-NS & BH-BH & Comments }
\startdata
A & 12-19  (0.92)\tablenotemark{b} & 0.07-0.11 (8.32) & 0.02-0.03 (71.1) & reference model \\
B & 7.6-12  (1.12) & 0.09-0.14 (8.27) & 0.01-0.02 (64.0) & full CE accretion \\
C & 68-101 (0.95) & 3.2-4.8   (7.25) & 7.7-11    (50.3) & CE for HG stars\\
\enddata
\label{rates}
\tablenotetext{a}{Range in rates results from different calibrations used.
Low rates are obtained with star formation calibration, while high rates are obtained 
with the supernova calibration.}
\tablenotetext{b}{In parenthesis we list distance scaling factors ${\cal
M}_{\rm dis}$, see eq.~\ref{dscale}.}
\end{deluxetable}

\begin{deluxetable}{lccc}
\tablewidth{300pt}
\tablecaption{LIGO Detection Rates\ \ [yr$^{-1}$]}
\tablehead{Model\tablenotemark{a} & NS-NS & BH-NS & BH-BH }
\startdata
A,LIGO & 0.003-0.005  & 0.0002-0.0003 & 0.0004-0.0006 \\
B,LIGO & 0.002-0.004  & 0.0002-0.0003 & 0.0003-0.0004 \\
C,LIGO & 0.017-0.025  & 0.0061-0.0090 & 0.10-0.15     \\
 & & & \\
A,AdLIGO & 13-19  & 0.68-1.1 & 1.6-2.5 \\
B,AdLIGO & 11-17  & 0.82-1.3 & 1.1-1.8 \\
C,AdLIGO & 73-109 & 26-39    & 439-655 \\
\enddata
\label{ligo}
\tablenotetext{a}{Models are presented for initial LIGO (range of
18.4 Mpc) and advanced LIGO (AdLIGO: range of 300 Mpc).}
\end{deluxetable}

\clearpage

\clearpage

\clearpage

\clearpage

\clearpage

\clearpage

\end{document}